\documentclass[twocolumn,prb,amsmath,amssymb,showpacs]{revtex4}

\usepackage{graphicx}% Include figure files
\usepackage{dcolumn}% Align table columns on decimal point
\usepackage{bm}% bold math

\begin{document}

\title{Metal-insulator transition in YH$_x$: scaling of
the sub-THz conductivity}

\author{I. G. Romijn}

\author{A. V. Pronin} \altaffiliation {Also at: Institute of General Physics,
Russian Academy of Sciences, 119991 Moscow, Russia}

\author{H. B. Brom}

\author{A. F. Th. Hoekstra}

\affiliation{Kamerlingh Onnes Laboratory, Leiden University, P.O.
Box 9504, 2300 RA Leiden, The Netherlands}

\date{July 7, 2004}

Accepted to PRB

\begin{abstract}
The established scaling laws of the conductivity with temperature
and doping are strong indications for the quantum nature of the
metal-insulator transition in YH$_x$. Here we report the first
results on the frequency scaling of the conductivity. Samples were
brought from the insulating to the metallic phase by carrier
doping via illumination. In the metallic phase, the sub-terahertz
conductivity coincides with the dc data. These results do not
agree with the simplest picture of a quantum-phase transition.
\end{abstract}


\maketitle

Yttrium hydride (YH$_{x}$, $2 \leq x \leq 3$) demonstrates a
remarkable transition of its electronic and optical properties
upon change of hydrogen concentration: a thin YH$_{x}$ film can be
continuously (and reversibly) brought from a shiny metal at $x=2$
to a transparent dielectric at $x \rightarrow 3$, simply by
changing pressure of the surrounding hydrogen gas.
\cite{Huiberts96, Huiberts97} Hoekstra et al.  showed that the
metal-insulator transition (MIT) could be neatly passed under
constant hydrogen pressure (and, consequently, constant $x$) by
change carrier doping via ultraviolet illumination at low
temperature.\cite{Hoekstra01} The effect of light doping persists
up to temperatures around 200 K. This light induced MIT is not
only intriguing for its spectacular visible effects, but also for
the underlying fundamental physics principles. Although the
precise mechanism is not known, \cite{mit} pronounced
electron-electron interactions are posited to lead to the opening
of a large optical gap of Hubbard type. \cite{Eder97, Ng97} If
experimentally confirmed as the causative agent, \cite{Kelly97,
Gelderen00} metal hydride films would be placed in the broad class
of highly correlated materials. Unlike nearly all Mott-Hubbard
systems, however, the metal-insulator transition in YH$_{x}$ is
continuous with minor effects of structural changes, and could
provide a rare window on critical behavior in the strong electron
interactions limit.

The established scaling laws of the conductivity $\sigma$ with
temperature $T$ and doping $n$, \cite{Hoekstra01,Hoekstra03} are
strong indications for the quantum nature of the metal-insulator
transition in YH$_x$.\cite{Sondhi97,Sachdev99} This scaling
picture holds up to surprisingly high temperatures (50 K) and the
product of the static $\nu$ and dynamical $z$ critical exponents,
$z\nu$ = 6.0 $\pm$ 0.5 is anomalously large.

To shed more light on the quantum nature of the MIT, the frequency
dependence of conductivity, $\sigma (\omega)$, will be very
informative. In an extensive frequency range $\omega$ and $1/T$
will influence $\sigma$ in a similar way, which will lead to
$\omega/T$-scaling behavior,\cite{Sondhi97,Sachdev99} and at
relatively low frequencies $\sigma (\omega)$ will decrease with
frequency in a Drude-like fashion. \cite{Sachdev99}

Here we report the first results on $\sigma(\omega)$ in the
(sub)terahertz regime. Terahertz spectroscopy provides an unique
opportunity because the frequency of the probing radiation is
comparable to the temperatures of interest (4 - 50 K), and can
easily be tuned. Samples were brought from the insulating to the
metallic phase by light doping. In the metallic phase, the
sub-terahertz conductivity as function of $T$ and $n$, is found to
display exactly the same power-low behavior as the dc
conductivity. We argue that these results do not agree with the
simplest predictions of a quantum-phase transition.

The  experimental setup for the sub-terahertz measurements is
described in detail in Ref. \onlinecite{Reedijk00}. We measured
the transmission through the yttrium hydride films on sapphire
substrates at three frequencies (275, 459, and 551 GHz) as a
function of temperature and for several doping levels on the metal
($n > n_{c}$) and insulator ($n < n_{c}$) sides of the MIT
($n_{c}$ is the critical electron concentration).  The samples
were prepared in the same way as the samples used by Hoekstra
\textit{et al.} \cite{Hoekstra01} As a substrate we used an
\textit{ac}-plane cut 0.5 mm thick sapphire platelet with a
diameter of 30 mm. Four gold contacts for van der Pauw dc
resistivity measurements were sputtered on top of it, followed by
a 300 nm yttrium layer. The latter was covered by a thin (20 nm)
layer of palladium. This layer is necessary since Pd acts as a
catalyst for the dissociation of the hydrogen gas molecules, and
increases the hydrogen sticking coefficient. The diameter of the Y
layer was 25 mm, and the diameter of the top Pd layer - 23 mm. In
the cryostat, the sample was put behind a 20-mm-diameter metal
diaphragm. Because the measurements were made in transmission, the
terahertz radiation was passing though all the layers. To estimate
the absorption in Pd layer, the transmission through a reference
sample with only palladium (without yttrium) was measured.

The yttrium films were loaded with 1 bar hydrogen during several
hours at room temperature in the same (modified) Oxford
Instruments optical cryostat, where the temperature scans were
performed. The continuously monitored dc resistivity showed that
yttrium was loaded to YH$_{2.9}$. \cite{dc_x} The YH$_{2.9}$
samples were brought into the metallic state by illumination
through the sapphire cryostat windows. As a light source an A5000
Xenon lamp (with power of about 1 kW) was used. During the
illumination, the sample space was kept below 10 K to prevent
heating of the sample.

\begin{figure}
\centering
\includegraphics[width=\columnwidth, clip]{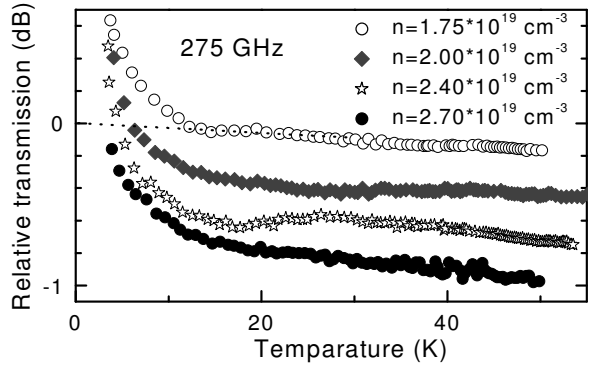}
\caption{Transmission through the yttrium hydride film on the
Al$_{2}$O$_{3}$ substrate with the additional layer of Pd on top,
measured at 275 GHz as a function of temperature. The zero of
relative transmission, which can be chosen arbitrarily, is here
set equal to the extrapolated value at $T = 0$ K (dotted line) of
the flat part of the lowest carrier concentration curve. }
\label{transmission1}
\end{figure}

\begin{figure}
\centering
\includegraphics[width=\columnwidth, clip]{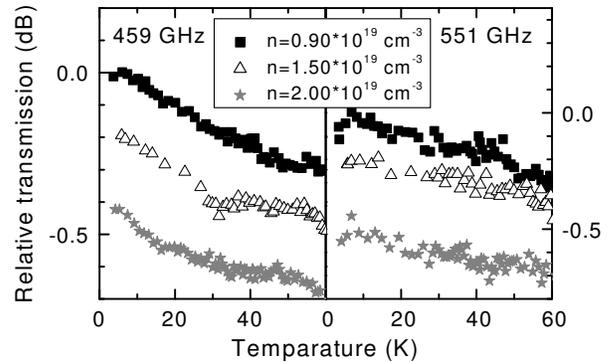}
\caption{Transmission through the same sample as in Fig. 1, but at
higher frequencies (459 and 551 GHz). Here the transmission at the
lowest carrier concentration and at T $\rightarrow$ 0 K is set as
0 dB. For better presentation, the vertical scales of two frames
are somewhat different.} \label{transmission2}
\end{figure}

Figs. \ref{transmission1} and \ref{transmission2} show the
experimentally measured (relative) transmission as a function of
temperature at 275, 459 and 551 GHz and several carrier
concentrations in YH$_{2.9}$. Only the changes between the
transmission values measured at different temperatures or at
different carrier concentrations are relevant. The carrier
concentrations have been determined from the dc resistivity
measurements by comparing our results with the data of Hoekstra
\textit{et al.} \cite{Hoekstra01} The correction for the
short-cutting Pd layer has been taken into account.
\cite{Romijn04} The data were collected during several hours of
slow heating from liquid helium temperature to 50 - 60 K, and then
averaged over 30 points. All data clearly demonstrate the
following tendency: the transmission decreases with increasing
temperature. That means the sub-terahertz conductivity goes up
with temperature at all carrier concentrations measured, which is
in qualitative agreement with the dc data of Refs.
\onlinecite{Hoekstra01} and \onlinecite{Hoekstra03}.

The transmission curves at 275 GHz (Fig. \ref{transmission1}) show
a steep increase at low temperatures ($T < 15$ K), while at higher
frequencies this effect is absent. Likely the increase is due to
diffraction of the longer wavelength radiation on the edges of the
sample diaphragm in combination with the multiple reflections
between the sample and the cryostat windows. For that reason the
data measured at 275 GHz below $\sim$ 15 K are not analyzed
further.

To find the absolute values of transmission (the transmission
coefficient), we performed frequency sweeps at a fixed temperature
(4.2 K). The standing wave pattern, which is always present in
(sub)terahertz measurements due to multiple reflections from all
the surfaces in the optical path like cryostat windows or lenses,
can be directly seen in this type of measurements, and the beam
path can be optimized to minimize the effect of the standing
waves. The remaining standing wave pattern can be eliminated by a
Fourier filtering. \cite{Reedijk00} In principle for an accurate
evaluation of the absolute values of the sub-terahertz
conductivity in YH$_{x}$, one should first determine the optical
parameters (complex index of refraction) of the bare sapphire
substrate as a function of frequency and temperature; then measure
the complex transmission (the amplitude and the phase shift) of
the Pd film on this substrate; calculate, applying Fresnel
formulas for a two-layer system, the optical constants (for
instance, the complex optical conductivity) of palladium; and,
finally, measure the complex transmission of the three-layer
system (sapphire-Pd-YH$_{x}$). From this complex transmission and
using the pre-defined optical constants of sapphire and Pd, one
can calculate the optical conductivity of the YH$_{x}$ films.

In our case, this procedure can be simplified. First, we found
that the optical properties of sapphire at the frequencies used
are not $T$ dependent below 50 K. Also the response of the Pd
layer can be very well describe by the Drude model in its
low-frequency limit (relaxation rate is much higher than the
measurement frequency). In addition, the conductivity (and
transmission) of the Pd layer does not show any temperature
dependence below 50 K (likely, due to the fact that the Pd film
contains a lot of defects and imperfections). Next, from the
Fresnel formulas, one can show that the transmission coefficient
$Tr$ (the transmission as compared to the empty optical channel)
of an optically thin film on a substrate at a \textit{fixed}
frequency, can be written as:
\begin{equation}
Tr = \frac{A}{1+ [(2\pi/c) \sigma d]^2}~, \label{Tr}
\end{equation}
here $A$ is the transmission of the substrate at this frequency,
$\sigma$ the conductivity of the film, $d$ its thickness, and $c$
the light velocity. By optically thin film we mean a film with an
optical thickness much less than the wavelength of the probing
radiation: $n_{f} \cdot d \ll \lambda$ and $k_{f} \cdot d \ll
\lambda$ ($n_{f}$ and $k_{f}$ are the real and the imaginary parts
of the complex refraction index of the film). These conditions
hold for our films: the minimal $\lambda$ we used was around 0.5
mm (frequency 551 GHz), while the YH$_{x}$ film thickness was 300
nm = $0.3 \times 10^{-3}$ mm, and the indexes $n_{f}$ and $k_{f}$
for YH$_{x}$ are certainly less than $10^2$ - the typical value
for good metals like Nb.

\begin{figure}[t]
\centering
\includegraphics[width=\columnwidth, clip]{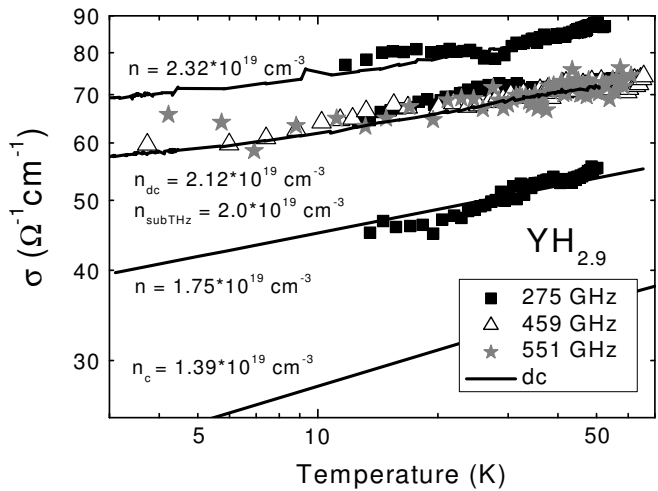}
\caption{Conductivity of YH$_{2.9}$ above the MIT. Symbols:
sub-THz data, lines: dc data from Refs. \onlinecite{Hoekstra01,
Hoekstra03} (the dc curve for $n = 1.75 \times 10^{19}$ cm$^{-3}$
have been calculated using the scaling low found in Refs.
\onlinecite{Hoekstra01} and \onlinecite{Hoekstra03}). The lowest
dc curve corresponds to the critical concentration $n_{c}$.}
\label{sigma}
\end{figure}

\begin{figure}[t]
\centering
\includegraphics[width=\columnwidth, clip]{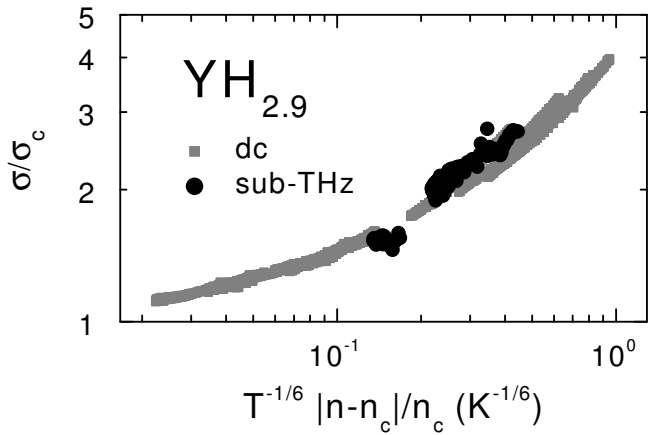}
\caption{Scaling of the sub-THz and dc conductivity in YH$_{2.9}$
above the MIT. The dc data are taken from Refs.
\onlinecite{Hoekstra01} and \onlinecite{Hoekstra03}. The sub-THz
data for 275, 475, and 551 GHz are recalculated from Fig.
\ref{sigma}. The critical conductivity $\sigma_{c}$ is defined as
described in Refs. \onlinecite{Hoekstra01} and
\onlinecite{Hoekstra03}, $n_{c} = 1.39 \times 10^{19}$ cm$^{-3}$.
The product of the critical exponents $z \nu = 6$ sets the power
of $T$ on the $x$-axes.} \label{scaling}
\end{figure}

If there is another absorbing layer on the substrate (as the Pd
layer in our case), the right part of Eq. \ref{Tr} should be
multiplied with a pre-factor, which takes into account absorbtion
in this layer. That simply leads to a renormalization of $A$.
Since we found that optical responses of the bare substrate and of
the Pd layer are not temperature dependent below 50 K, the only
temperature dependent parameters in Eq. \ref{Tr} are the
experimentally measured transmission $Tr$ and the conductivity
$\sigma$. Thus, by inverting Eq. \ref{Tr}:
\begin{equation}
\sigma = \frac{c}{2\pi d} \sqrt{A/Tr-1}~, \label{Sg}
\end{equation}
one can calculate the temperature dependence of the
yttrium-hydride conductivity. The accuracy in determination of $A$
(10\%) and of the absolute transmission determines the accuracy of
conductivity. For $n > 1.7 \times 10^{19}$ cm$^{-3}$, the
difference between $A$ and the absolute transmission $Tr$ is
sufficient to calculate $\sigma$ accurately. The results are shown
in Fig. \ref{sigma}. We found that at these concentration levels,
the absolute values of the sub-THz conductivity coincide with the
dc data at all temperatures measured.

Thus, on the metallic side of the phase transition, the
high-frequency conductivity follows the same scaling law as the dc
conductivity. The collapsed data are shown in Fig. \ref{scaling}
as function of $T^{-1/z \nu} |n-n_{c}| / n_{c}$, $z \nu = 6$.
Apparently the product of the critical exponents $z \nu$ remains
the same.

The frequencies at which the data were collected (275, 459, and
551 GHz) may be translated into the temperature units: 13.2, 22.0,
and 26.4 K. Our temperature range is 4 - 50 K. Thus, from a formal
point of view, an $\omega/T$-scaling is expected since the sample
is in the intermediate regime ($k_{B}T \approx \hbar \omega$). But
we see no "real" $\omega/T$-scaling: the conductivity for all
frequencies follows the dc behavior: $\sigma(\omega,T) =
\textsl{F}(T^{-1/6} |n-n_{c}| / n_{c})$ with no frequency
dependence. Apparently, also no Drude-like frequency dependence is
needed to collapse the sub-THz data onto the dc
curves.\cite{Sachdev99} It is clear that these results do not
agree with the simplest picture of a quantum-phase transition.

Summarizing, we performed measurements of frequency-dependent
conductivity in YH$_x$ ($x \approx 2.9$) at sub-THz frequencies
and at several carrier concentrations. The sample was brought from
the insulating to the metallic phase by light doping. In the
metallic phase, contrary to theoretical expectations the
sub-terahertz conductivity as function of temperature and carrier
concentration, is found to be equal to the dc conductivity within
the experimental uncertainty. In the insulating phase the accuracy
of the data are too strongly influenced by the value of $A$. For
that reason we have redesigned the set-up and new measurements are
planned in the immediate future.

We thank Ruud Westerwaal for assistance in sample preparation.
Fruitful discussions with Sahana Mungila, Wim van Saarloos, and
Jan Zaanen are greatly acknowledged. The work was supported by the
Dutch Foundation for Fundamental Research on Matter (FOM/NWO).


\begin{thebibliography}{10}
\bibitem{dc_x}
{The connection between the dc conductivity and the doping level
$x$ has been established in Refs. \onlinecite{Huiberts97,
Hoekstra01}}.
\end{thebibliography}
\end{document}